\documentclass[12pt,notitlepage,a4paper]{article}
\pdfoutput=1
\usepackage[a4paper,margin=1in]{geometry}
\usepackage{amsmath,amssymb,amscd,amsfonts}
\usepackage{mathtools}
\numberwithin{equation}{section}
\usepackage{url}
\usepackage[shortlabels]{enumitem}
\usepackage{tikz} 
\usetikzlibrary{decorations}
\usetikzlibrary{arrows,decorations.markings}         
\usepackage[hidelinks]{hyperref}
\usepackage[open,
  openlevel=2,
  atend,
  numbered]{bookmark}
\usepackage{graphicx}
\usepackage[final]{pdfpages}
\usepackage{threeparttable}
\usepackage{caption}
\usepackage{float}
\usepackage{stackengine}
\usepackage{scalerel}
\usepackage[bf]{titlesec}
\titlespacing{\section}{3pc}{2pc}{0.8pc}
\titleformat{\section}
  {\centering\large\bfseries} 
  {\Large\thesection}
  {0.5em} 
  {\Large}  
\titlespacing{\subsection}{0pc}{2pc}{0.8pc}
\titleformat{\subsection}
  {\centering\bfseries} 
  {\normalsize\thesubsection}
  {0.5em} 
  {\normalsize}  
\usepackage{tocloft}

\newlength{\mylen}	
\urldef\footurlb\url{cocalc.com/dfriedan/DM/SM}
\urldef\footurla\url{physics.rutgers.edu/~friedan}
\def\eq{\begin{equation}}
\def\en{\end{equation}}
\def\eqg{\eq\begin{gathered}}
\def\eng{\end{gathered}\en}
\def\eqa{\eq\begin{aligned}}
\def\ena{\end{aligned}\en}

\def\EW{\scriptscriptstyle\mathrm{EW}}

\def\Higgs{\scriptscriptstyle \mathrm{Higgs}}

\def\CGF{\scriptscriptstyle \mathrm{CGF}}

\def\sunit{\text{\footnotesize s}}

\def\ntimes{\,{\times}\,}

\stackMath
\def\dyhat{-0.2ex}
\newcommand\myhat[1]{\ThisStyle{%
              \stackon[\dyhat]{\SavedStyle#1}
                              {\SavedStyle\hat{\phantom{#1}}}}}
\def\that{\kern0.1em\myhat{\kern-0.1em t}}

\def\kg{\mathrm{kg}}
\def\munit{\mathrm{m}}
\def\cmunit{\text{cm}}
\def\Msun{\textup{M}_{\odot}}
\def\texttilde{\raise-0.7ex\hbox{\!\texttt{\char`\~}}}
\def\rhob{\rho_{b}}
\def\rb{r_{b}}
\def\mb{m_{b}}
\def\tot{\mathrm{tot}}

\titlespacing{\section}{3pc}{2pc}{0.8pc}
\titleformat{\section}
  {\centering\normalsize\bfseries} 
  {\large\thesection}
  {0.5em} 
  {\large}  
\begin{document}
\def\title{Dark matter stars}
\begin{center}
{\LARGE \title}
\vskip4ex
{\large Daniel Friedan}
\vskip2ex
{\it
New High Energy Theory Center
and Department of Physics and Astronomy,\\
Rutgers, The State University of New Jersey,\\
Piscataway, New Jersey 08854-8019 U.S.A. and
\vskip1ex
Science Institute, The University of Iceland,
Reykjavik, Iceland
\vskip1ex
\href{mailto:dfriedan@gmail.com}{dfriedan@gmail.com}
\qquad
\href{https://physics.rutgers.edu/\textasciitilde friedan/}
{physics.rutgers.edu/\texttilde friedan}
}
\vskip2ex
March 22, 2022
\end{center}
%
%
%
\begin{center}
\vskip3ex
{\sc Abstract}
\vskip2.5ex
\parbox{0.96\linewidth}{
\hspace*{1.5em}
The dark matter in the CGF cosmology is a cosmological SU(2)-weak 
gauge field (the CGF).
The TOV stellar structure equations are solved numerically for
stars composed of this dark matter.
The star mass $M$ can take any value up to
a maximum $9.14 \ntimes 10^{-6}\,\Msun$.
For each value of $M$ the star radius $R$ 
lies between $5.23\, \cmunit$ and $13.6\,\cmunit$.
More than one value of $R$ is possible
when $M> 5.09 \ntimes 10^{-6}\,\Msun$.
For those stars,
a transition from larger to smaller $R$ would
release gravitational energy on the order of $10^{41}\,\mathrm{J}$
in a time on the order of $10^{-10}\,\mathrm{s}$.
}
\end{center}

\vspace*{-2ex}

%
%
%
\begin{center}
\tableofcontents
\end{center}
%
%

\section{Introduction}
The CGF cosmology is a complete theory
of the Standard Model cosmological epoch,
from the electroweak transition onward
\cite{Friedan:2020poe,Friedan2022:Atheory}.
The theory has no free parameters 
and  assumes no physical laws beyond the Standard Model and General Relativity.
All of cosmology is given by the time evolution of
a uniquely determined highly symmetric semi-classical initial state
in the period leading up to the electroweak transition.

The CGF universe in the leading order, classical approximation contains only dark matter, no ordinary matter.
The dark matter is a cosmological SU(2)-weak gauge field (the CGF).
The relatively small amount of ordinary matter in the universe is a 
next to leading order correction to the dark matter universe
from the small fluctuations of the Standard Model 
fields around the classical CGF.

The CGF behaves effectively as a perfect fluid.  Its  equation of state
was derived in \cite{Friedan2022:Atheory} in the leading order, classical approximation.
Initial fluctuations of the CGF
presumably collapsed gravitationally to form self-gravitating objects.
Here,
the Tolman-Oppenheimer-Volkoff stellar structure equations
for stars composed of CGF dark matter
are solved numerically.
The results are presented
and some features are noted.

The numerical calculations are done in SageMath \cite{sagemath9.4}
using the mpmath arbitrary-precision floating-point arithmetic 
library \cite{mpmath}.
The Sagemath notebooks along with printouts of the notebooks
are provided in the Supplemental Materials \cite{Friedan2022:DMstarsSuppMat}.

\section{Tolman-Oppenheimer-Volkoff equations}
The scale of the CGF fluid is set by the density (in $c=1$ units)
\eq
\rhob = \frac{m_{\Higgs}^{4}}{\hbar^{3}}
= 5.68 \ntimes 10^{28}\, {\kg}/{\munit^{3}}
\en
which is $10^{11}$ times larger than the density of a neutron star.
$m_{\Higgs}$ is the mass of the Higgs boson.
The associated gravitational distance and mass scales are
\eq
\begin{aligned}
\rb  &= (4\pi G \rhob)^{-1/2}
 = 1.45 \ntimes 10^{-10} \,\sunit
 = 4.34 \,\cmunit
\\[1ex]
\mb &= G^{-1} \rb 
= 5.85\ntimes 10^{25} \,\kg
= 2.94\ntimes 10^{-5} \Msun
= 5.26 \ntimes 10^{42}\,\text{J}
\end{aligned}
\en

The TOV stellar structure equations describe the
spherically symmetric  static
configurations of a self-gravitating perfect fluid
such as the CGF.
The TOV equations in dimensionless variables are
\eqg
\frac{d\hat p}{d\hat r} = - \frac{(\hat \rho+\hat p) 
( \hat m +\hat r^{3}\hat p)}
{\hat r(\hat r-2\hat m)}
\qquad
\frac{d\hat m}{d\hat r} = \hat r^{2}  \hat \rho\hat
\qquad
\frac{d \hat e}{d\hat r} = 
\hat r^{2} \hat \rho  \left(1-\frac{2\hat m}{\hat r}  \right)^{-1/2}
\\[2ex]
\hat r = \frac{r}{\rb}
\qquad
\hat m = \frac{m}{\mb}
\qquad
\hat e = \frac{E_{\tot}}{m_{b}}
\qquad
\hat \rho = \frac\rho{\rhob}
\qquad
\hat p = \frac{p}{\rhob}
\eng
$\hat r$ is the radial distance,
$\hat m(\hat r)$ and
$\hat e(\hat r)$ are the mass and the total energy inside $\hat r$,
$\hat \rho(\hat r)$ 
and $\hat p(\hat r)$  are the density and pressure at $\hat r$.
Given an equation of state relating $\hat p$ and $\hat \rho$,
the TOV equations can be integrated starting from $\hat r=0$ with initial 
condition the central density $\hat \rho(0)$.

\section{CGF equation of state}

\begin{figure}[H]
\begin{center}
\includegraphics[scale=.8]{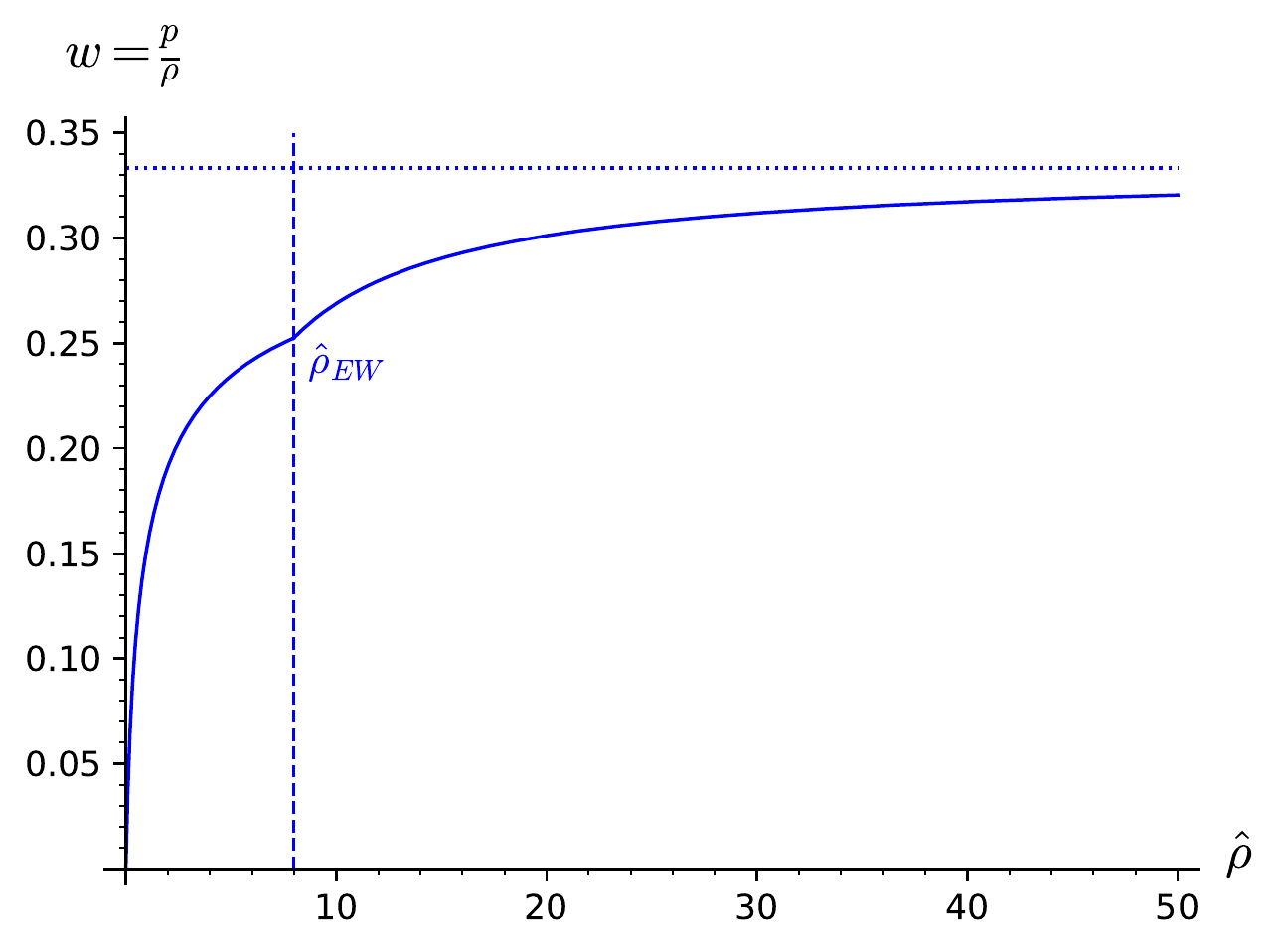}
\caption{\label{fig:w}}
\end{center}
\end{figure}

The CGF equation of state derived in \cite{Friedan2022:Atheory} is
\eq
{ \everymath={\displaystyle}
\begin{array}{cll}
\hat \rho \ge \hat \rho_{\EW}\colon
 &
\hat p =\frac13 \left(\hat \rho  -  c_{b}\, \hat\rho_{\EW}\right)
\quad&
\hat\rho_{\EW}= 7.97
\quad
c_{b} = 0.243
\\[1ex]
\hat \rho \le \hat \rho_{\EW}\colon
&
\hat \rho,\,\hat p = 
\hat \rho_{\CGF}(k^{2}),\,\hat p_{\CGF}(k^{2})
\quad&
0 \le k^{2} \le \frac12
\end{array}
}
\en
At densities $\hat \rho \ge \hat\rho_{\EW}$
the CGF fluid is a simple mixture of radiation
and vacuum energy.
The numbers $\hat\rho_{\EW}$ and $c_{b}$ are algebraic expressions in the Higgs 
coupling constant $\lambda$ and the SU(2) gauge coupling constant $g$.
For $\hat \rho \le \hat \rho_{\EW}$ the equation of state is defined implicitly
by analytic functions $\hat \rho_{\CGF}(k^{2})$ and $\hat p_{\CGF}(k^{2})$ 
of the parameter $k^{2}$. 
The precise form of the two analytic functions is not illuminating.
They are algebraic expressions in $k^{2}$, $\lambda$, $g$, and the 
complete elliptic integrals $K(k)$ and $E(k)$.
The density $\hat \rho_{\CGF}(k^{2}) $ decreases monotonically from
$\hat\rho_{\EW}$ to $0$
as the parameter $k^{2}$ decreases from $1/2$  to $0$.
The equation of state simplifies in the limit $\hat\rho\rightarrow 0$ to
\eq
\hat p = \frac{c_{a}}2  \hat \rho^{2} + O(\hat \rho^{3})
\qquad
c_{a} = 0.992
\en
$c_{a}$ is another algebraic expression in $\lambda$ and $g$.
Figure~\ref{fig:w} plots the equation of state.

\section{The solutions}

\begin{figure}[H]
\begin{center}
\includegraphics[scale=0.52]{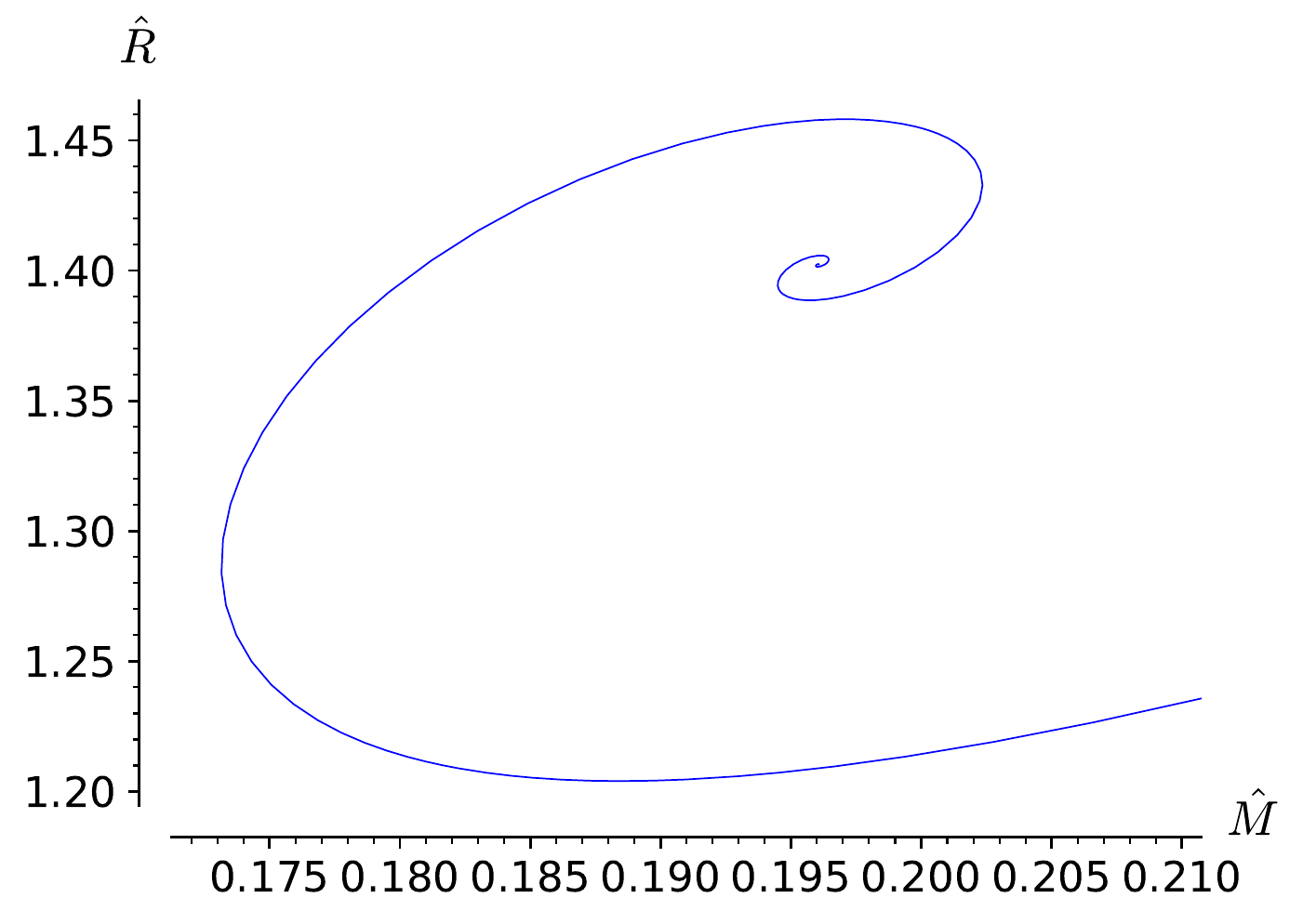}
\hfill
\includegraphics[scale=0.52]{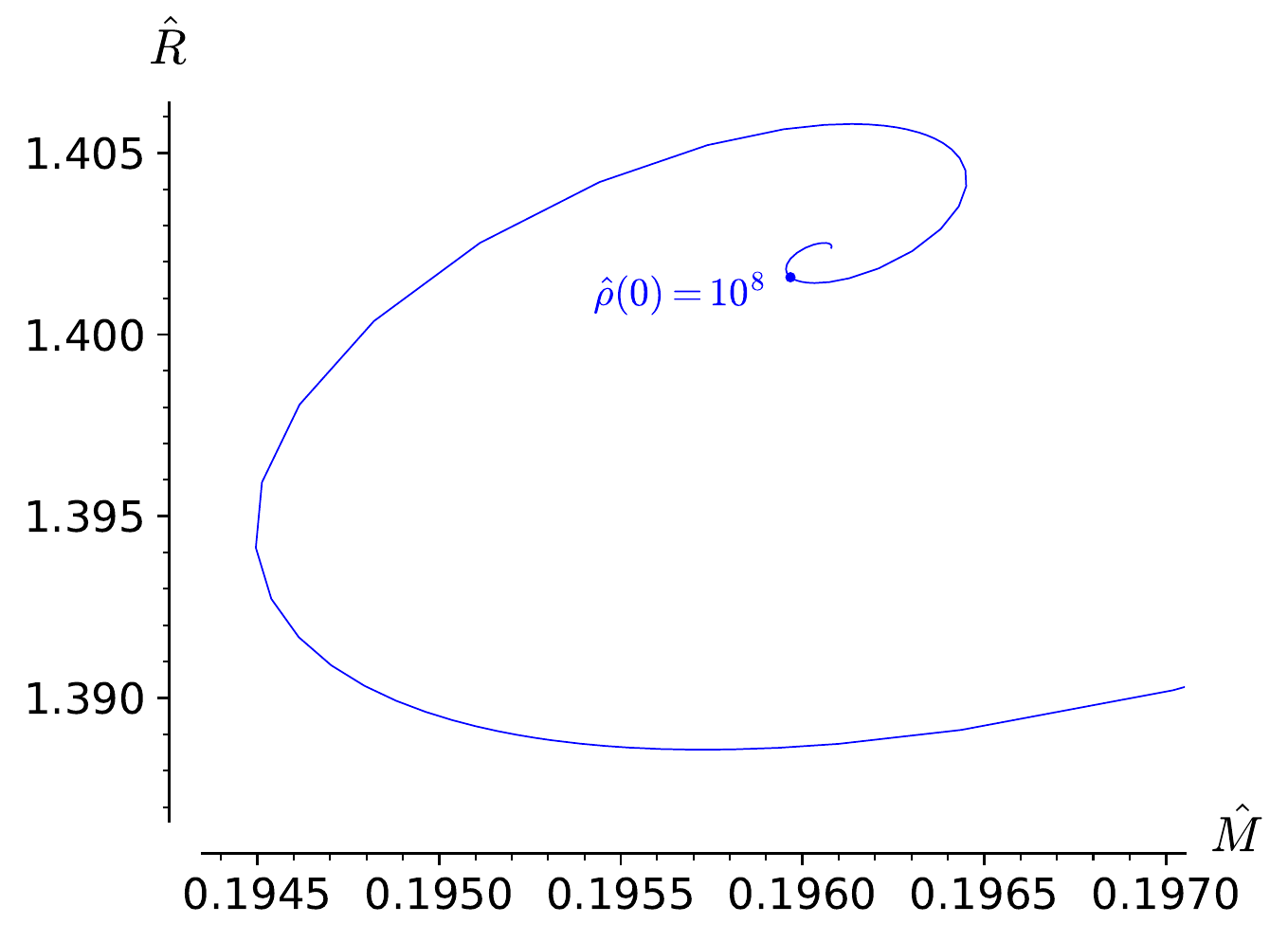}
\caption{\label{fig:MRblowup}}
\end{center}
\end{figure}

The TOV equations are solved numerically.
The calculations are shown in 
the Supplemental Materials \cite{Friedan2022:DMstarsSuppMat}.
For every
central density $\hat \rho(0) > 0$
the solution
reaches $\hat \rho=\hat p=0$ 
at a finite radial distance $\hat r=\hat R$.
So each star has a well-defined radius $\hat R$, 
a well-defined mass $\hat M = \hat m(\hat R)$, and 
a well-defined total energy $\hat E= \hat e (\hat R)$.
\begin{figure}
\begin{center}
\includegraphics[scale=0.9]{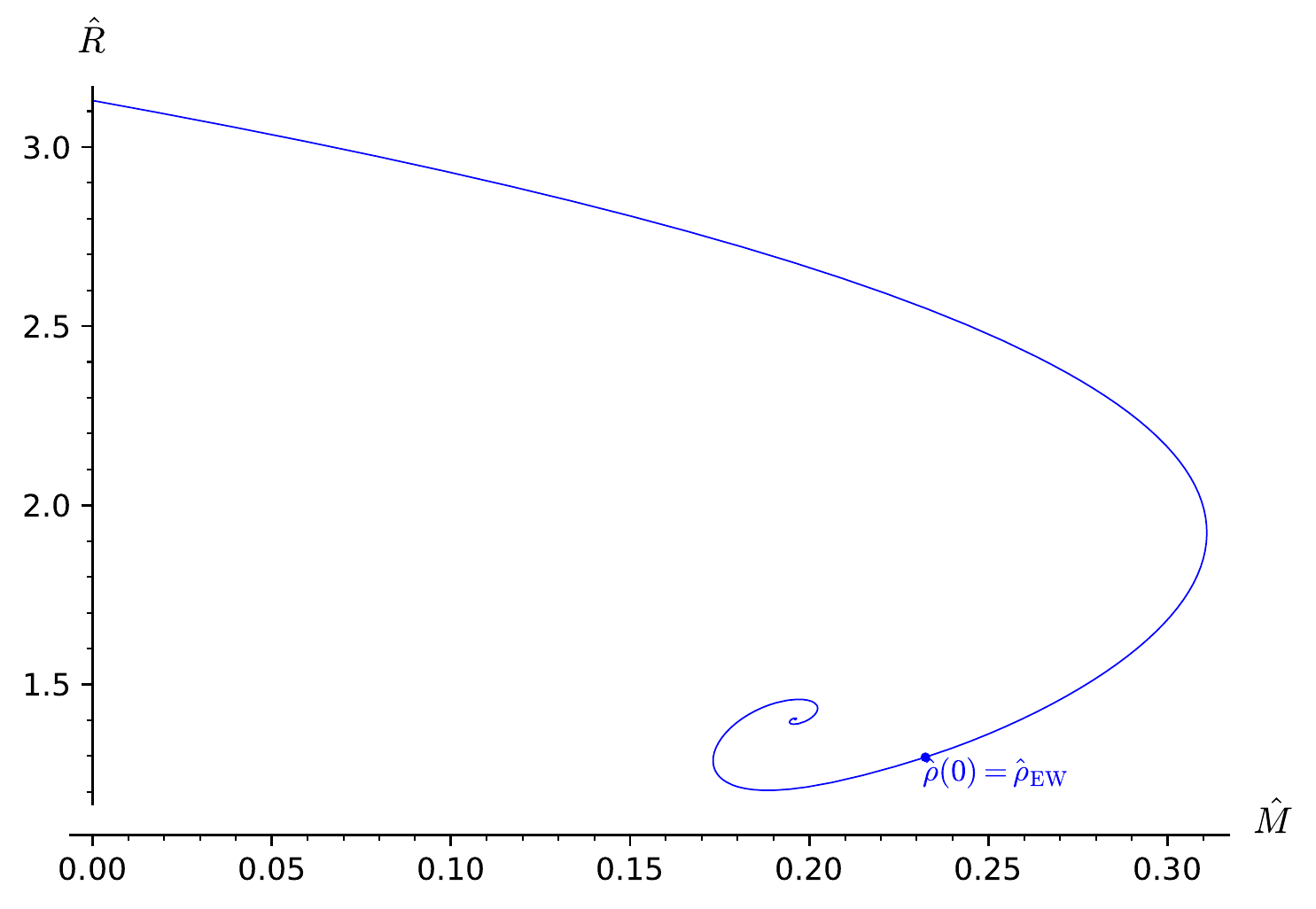}
\caption{\label{fig:MR}}
\end{center}
\end{figure}
Figure~\ref{fig:MR}
shows the stars  plotted in the $\hat M,\hat R$ plane.
The curve of stars is parametrized by $\hat \rho(0)$,
starting at $\hat \rho(0)=0$, $M=0$.
The curve spirals inward with increasing central density.
The plots in Figure~\ref{fig:MRblowup} are successive blowups showing the
curve spiraling towards a fixed point.
This asymptotic behavior is explained by the TOV 
equations for a pure radiation fluid ($w=1/3$).
When the CGF central density is large,
most of the evolution in $\hat r$ 
takes place with equation of state very close to that of pure radiation.  The TOV 
equations for pure radiation are the flow equations of a vector field in 
the plane with a fixed point \cite{Friedan2022:DMstarsSuppMat}.

In Figure~\ref{fig:MR} the labeled point on the curve 
is the star with $\hat \rho(0) = \hat \rho_{\EW}$.
At central densities larger than $\hat\rho_{\EW}$, i.e.~inwards along the spiral,
the star consists of a core satisfying the high density equation of state 
and an outer shell satisfying the low density equation of state.
Within the core, the CGF holds the Higgs field at $\phi=0$.
In the shell, $\phi^{\dagger}\phi$ increases from 0 at the core-shell 
boundary to the vacuum 
expectation value $v^{2}/2$ at the stellar surface.
Outwards on the spiral from $\hat\rho_{\EW}$,
at central densities $\hat\rho(0)\le \hat\rho_{\EW}$,
the stars are coreless.

\section{Abundance, clouds, seeds, binding energy}

The minimum radius is
$\hat R = 1.20$, $R = 5.23\,\cmunit$.
The maximum radius is
$\hat R = 3.13$, $R = 13.6\,\cmunit$.
The maximum star mass is $\hat M = 0.311$,
$M = 1.82 \ntimes 10^{25}\,\kg=9.14\ntimes 10^{-6}\Msun$.
The abundance distribution
on the star curve is calculable from first principles in the CGF 
cosmology,
but that is beyond the scope of this paper.
Microlensing observations put an upper limit 
$M< 10^{-11}\Msun$
on compact objects 
as the halo dark matter \cite{Niikura:2017zjd}.
Therefore most of the abundance distribution will have to be concentrated at
small mass on the dark matter star curve,
the limit $\hat \rho(0)\rightarrow 0$.

The equation of state in the $\hat \rho(0)\rightarrow 0$
limit is $\hat p = \frac12 c_{a} \hat\rho^{2}$.
The TOV equations are non-relativistic at leading order in $\hat \rho(0)$
and can be solved.
\eqg
\hat r = c_{a}^{1/2}\theta
\qquad
\hat \rho(\hat r) =   \frac{\sin\theta}{\theta}\hat\rho(0)
\qquad
\hat m(\hat r) =  c_{a}^{3/2} \left(\sin \theta
- \theta \cos \theta\right) \hat\rho(0)
\\[1ex]
\hat R = \pi c_{a}^{1/2} = 3.13
\qquad
\hat M =  \pi c_{a}^{3/2} \hat\rho(0) = 3.11 \,\hat\rho(0)
\\[1ex]
\hat E - \hat M =\frac{3\pi c_{a}^{5/2}}{4}  \hat\rho(0)^{2}
= 2.31 \, \hat\rho(0)^{2}
\eng
All the small mass stars
have almost exactly the same radius $\hat R = 3.13$,
$R=13.6\, \cmunit$.

We might speculate that the initial CGF fluctuations condense gravitationally to an ensemble 
of dark matter stars.
A small fraction of the total mass is in massive dark matter stars.
These become seeds for ordinary stars.
The halos become populated by low mass dark matter stars --- 
geometrically 
identical  spheres of dark matter
at low density.

For mass $\hat M > 0.173$,
$M > 5.09 \ntimes 10^{-6}\Msun$
more than one 
value of the radius $\hat R$ is possible.
Suppose that one of the solutions at mass $\hat M$ is stable and the 
others metastable.
A star at a metastable radius might be provoked
by an external influence to drop to a radius at lower gravitational energy,
releasing the difference in energy.
If massive dark matter stars served as seeds for 
formation of ordinary stars, some metastable dark matter stars
might lurk in the centers of
ordinary stars until provoked to release such an outburst of gravitational energy.
Metastable dark matter stars might also wander without ordinary 
matter, emitting a burst of energy when provoked, say by a near 
collision.

\begin{figure}
\begin{center}
\includegraphics[scale=0.48]{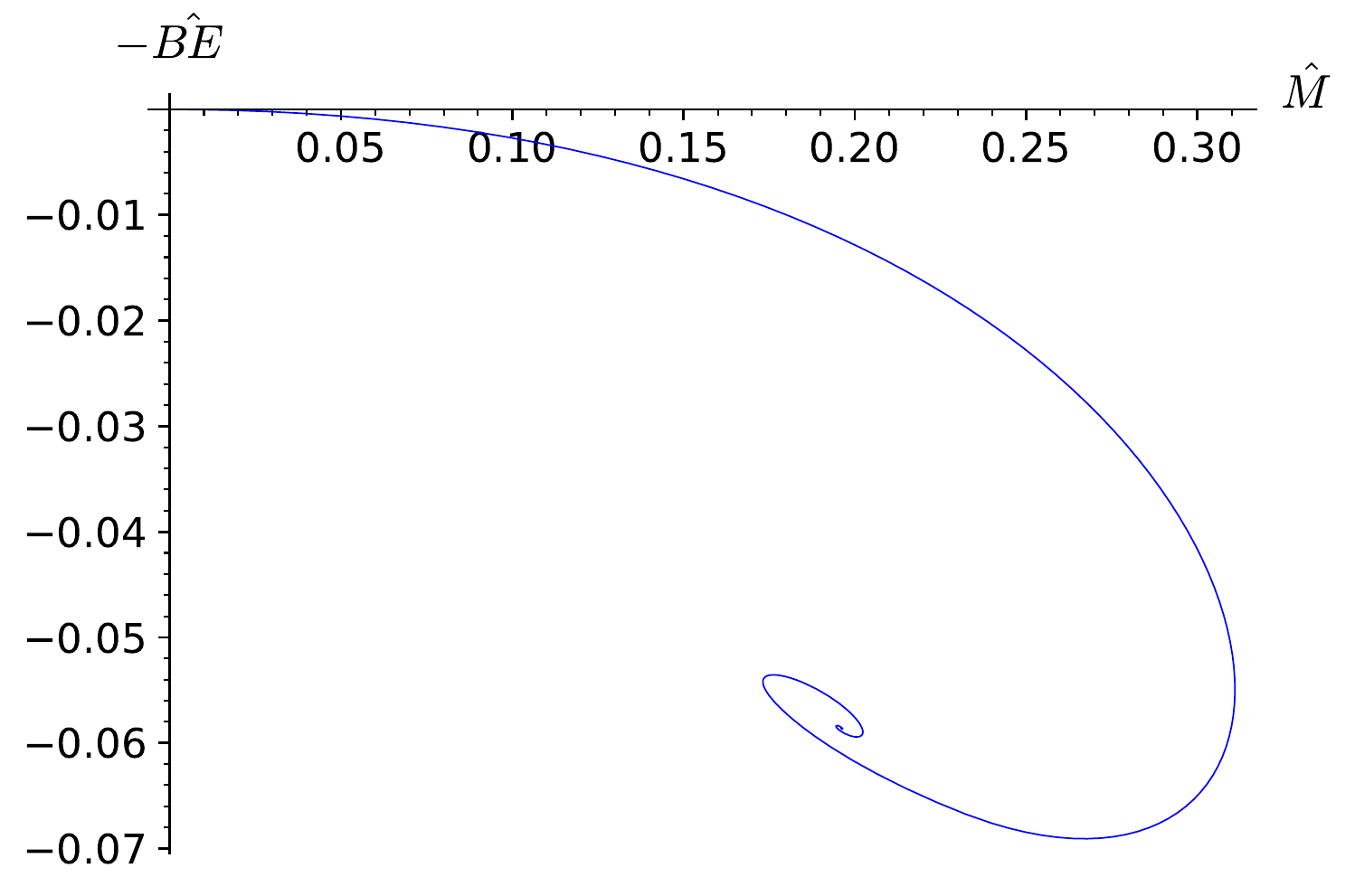}
\hfill
\includegraphics[scale=0.48]{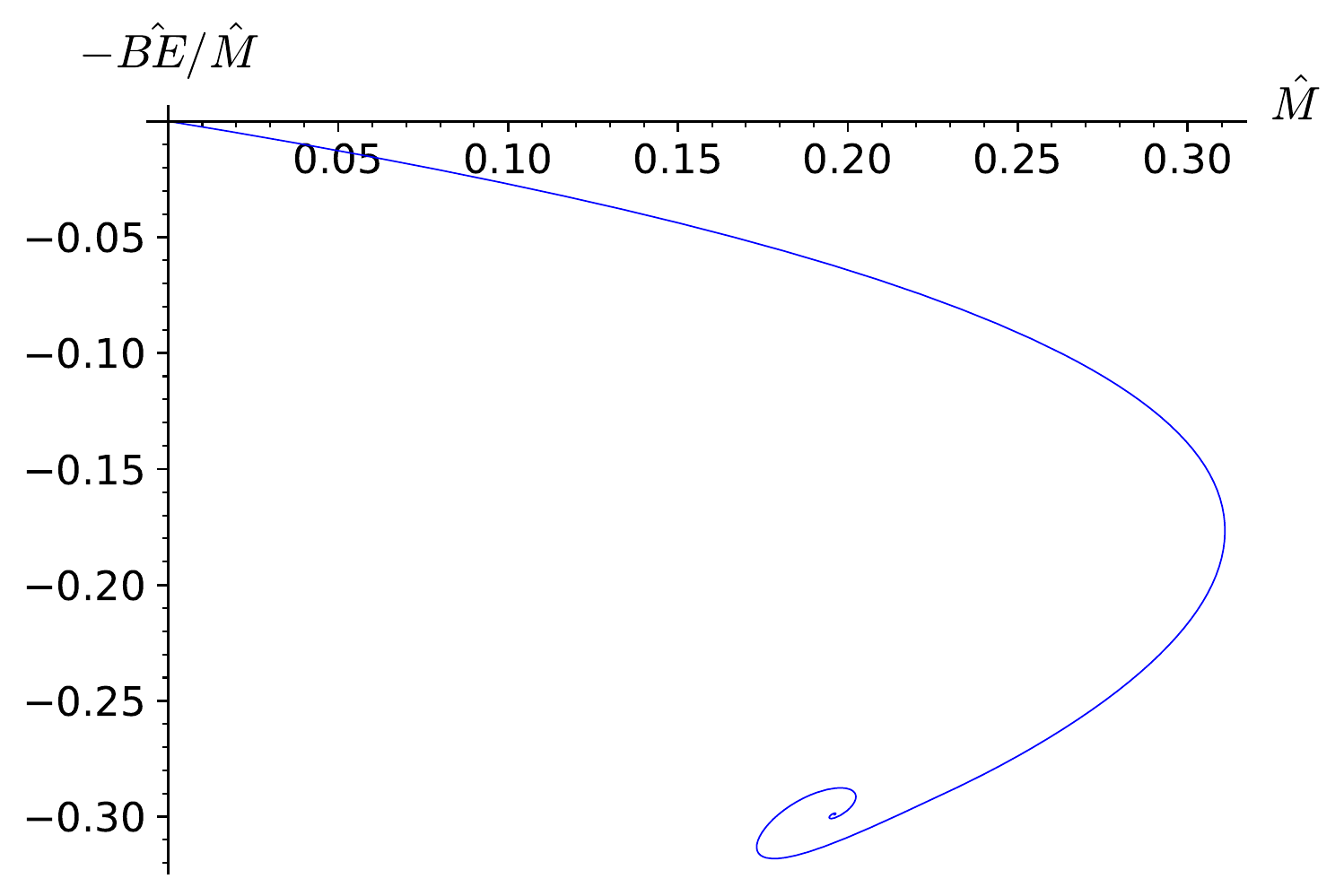}
\caption{\label{fig:BE}}
\end{center}
\end{figure}

The duration of an outburst from a metastable dark matter star
will be on the order of the star radius,  $10^{-10}\,\sunit$.
To get a handle on the outburst energy, Figure~\ref{fig:BE}
shows the binding energy $\hat{\mathrm{BE}} = \hat E - \hat M$ 
and the ratio ${\mathrm{BE}}/M = \hat{\mathrm{BE}}/\hat M$
(both plotted so that going downwards is energetically favored).
The curve of binding energies shows that the lowest energy state for 
given $\hat M$ is the state of lowest radius.
The drops in binding energy range up to about $0.05\,m_{b} = 
10^{41}\,\mathrm{J}$.
This is about $10^{-3}$ of the energy released in a 
supernova.

The slope of the binding energy curve shows that fusion of low mass stars
with stars of almost any mass is energetically favorable.
Bubbling off a small mass star is not favored.
The curve of binding energy/mass shows that 
fission of high mass stars is not favored.
Ultimately (on some time scale)
all the stars will end at the bottom of the $\hat{\mathrm{BE}}/\hat 
M$ curve.

\section{Dark matter stars probing energies beyond the Standard Model}

The energy scale at the center of the star is
$\hat \rho(0)^{1/4} m_{\Higgs}$
so high central density stars
probe energies beyond the standard model.
The marked star 
in Figure~\ref{fig:MRblowup}
is at central density $\hat \rho(0)=10^{8}$.
The energy scale at the star center is $10\,\mathrm{TeV}$.
All the stars inward from there along the spiral curve
probe energies $> 10\,\mathrm{TeV}$.
They would be metastable,
with a rich set of possible transitions.


\phantomsection
\section*{Acknowledgments}
\addcontentsline{toc}{section}{\numberline{}Acknowledgments}
I thank C.~Keeton 
for advice on microlensing and for suggesting reference \cite{Niikura:2017zjd}. 
This work was supported by the Rutgers New High Energy Theory Center
and by the generosity of B. Weeks.
I am grateful
to the  Mathematics Division of the 
Science Institute of the University of Iceland
for its hospitality.

\bibliographystyle{ytphys}
\raggedright
\phantomsection
\addcontentsline{toc}{section}{\numberline{}References}
\bibliography{Literature}

\end{document}